\documentclass[sigconf]{acmart} 

\acmConference[ICSE 2026]{48th International Conference on Software Engineering}{April 2026}{Rio de Janeiro, Brazil}

\copyrightyear{2026}
\acmYear{2026}
\setcopyright{cc}
\acmConference[ICSE '26]{2026 IEEE/ACM 48th International Conference on Software 
Engineering}{April 12--18, 2026}{Rio de Janeiro, Brazil}
\acmBooktitle{2026 IEEE/ACM 48th International Conference on Software 
Engineering (ICSE '26), April 12--18, 2026, Rio de Janeiro, Brazil}
\acmPrice{}
\acmDOI{10.1145/3744916.3764527}
\acmISBN{979-8-4007-2025-3/26/04}

\ccsdesc[500]{Software and its engineering~Software defect analysis}
\ccsdesc[300]{Software and its engineering}

\newcommand{\code}[1]{\texttt{\small#1}} 
\newcommand{\scode}[1]{\texttt{\footnotesize#1}} 

\usepackage{enumitem}
\setlist{leftmargin=6mm}
\usepackage{booktabs}
\usepackage{url}
\usepackage{amsthm}
\theoremstyle{definition}

\usepackage{amsmath}
\usepackage{booktabs}
\usepackage{graphicx}
\usepackage{multirow}
\usepackage{xcolor}
\usepackage{listings}
\definecolor{halfgray}{gray}{0.35}
\definecolor{deepblue}{rgb}{0,0,0.5}
\definecolor{deepred}{rgb}{0.6,0,0}
\definecolor{deepgreen}{rgb}{0,0.5,0}
\definecolor{highlightorange}{rgb}{0.98, 0.92, 0.8}

\makeatletter
\newcommand\HL{%
	\gdef\lst@alloverstyle##1{%
		\fboxrule=0pt
		\fboxsep=0pt
		\colorbox{lightgray}{\strut##1}%
	}%
}
\newcommand\HLoff{%
	\xdef\lst@alloverstyle##1{##1}%
}

\definecolor{lightgreen}{rgb}{0.9,1,0.9}
\definecolor{lightred}{rgb}{1,0.9,0.9}

\makeatletter
\newcommand\HLgreen{%
	\gdef\lst@alloverstyle##1{%
		\fboxrule=0pt
		\fboxsep=0pt
		\colorbox{lightgreen}{\strut##1}%
	}%
}
\newcommand\HLgreenoff{%
	\xdef\lst@alloverstyle##1{##1}%
}

\makeatletter
\newcommand\HLred{%
	\gdef\lst@alloverstyle##1{%
		\fboxrule=0pt
		\fboxsep=0pt
		\colorbox{lightred}{\strut##1}%
	}%
}
\newcommand\HLredoff{%
	\xdef\lst@alloverstyle##1{##1}%
}

\newcommand{\name}{Testora}
\newcommand{\anonID}[1]{#1}

\newcommand{\nbRegressions}{19}
\newcommand{\nbCoincidentalFixes}{11}

\newcommand{\precision}{55\%}
\newcommand{\recall}{67\%}

\newcommand{\dollarsPerPR}{0.003}
\newcommand{\inputTokensPerPR}{5,818}
\newcommand{\outputTokensPerPR}{3,622}
\newcommand{\totalTokensPerPR}{9,440}
\newcommand{\minutesPerPR}{12.3}

\begin{document}

\title[\name{}: Using Natural Language Intent to Detect Behavioral Regressions]{\name{}: Using Natural Language Intent to\\ Detect Behavioral Regressions}

\author{Michael Pradel}
\email{michael@binaervarianz.de}
\orcid{0000-0003-1623-498X}
\affiliation{%
  \institution{CISPA Helmholtz Center for Information Security}
  \city{Stuttgart}
  \country{Germany}
}

\begin{abstract}
As software is evolving, code changes can introduce regression bugs or affect the behavior in other unintended ways.
Traditional regression test generation is impractical for detecting unintended behavioral changes, because it reports all behavioral differences as potential regressions.
However, most code changes are intended to change the behavior in some way, e.g., to fix a bug or to add a new feature.
This paper presents \name{}, the first automated approach that detects regressions by comparing the intentions of a code change against behavioral differences caused by the code change.
Given a pull request (PR), \name{} queries an LLM to generate tests that exercise the modified code, compares the behavior of the original and modified code, and classifies any behavioral differences as intended or unintended.
For the classification, we present an LLM-based technique that leverages the natural language information associated with the PR, such as the title, description, and commit messages -- effectively using the natural language intent to detect behavioral regressions.
Applying \name{} to PRs of complex and popular Python projects, we find \nbRegressions{} regression bugs and \nbCoincidentalFixes{} PRs that, despite having another intention, coincidentally fix a bug.
Out of 13 regressions reported to the developers, 11 have been confirmed and 9 have already been fixed.
The costs of using \name{} are acceptable for real-world deployment, with \minutesPerPR{} minutes to check a PR and LLM costs of only \$\dollarsPerPR{} per PR.
We envision our approach to be used before or shortly after a code change gets merged into a code base, providing a way to early on detect regressions that are not caught by traditional approaches.
\end{abstract}

\keywords{Regression testing, test oracle, natural language processing, large language models, software maintenance}

\maketitle

\section{Introduction}

Practically all successful software projects are continuously evolving.
While changing code is necessary to fix bugs, add new features, improve performance, or increase the maintainability of a code base, code changes may also negatively impact the software.
For example, while trying to fix a specific bug, a developer may not only remove the buggy behavior, but also accidentally introduce a new bug or modify the behavior in some other unintended way.
Prior work shows that behavioral changes are common in practice, leading to regression bugs that often remain unnoticed~\cite{mostafa2017experience}.

As a real-world example of a regression bug, consider Figure~\ref{fig:example}, which shows a pull request (PR) of the popular Python \emph{scipy} library.
The title and description of the PR indicate that the change is intended to add array API support to the \code{differential\_entropy} function.
Adding array API support is a larger design change in scipy, which affects how the library internally handles different array types, but it should not change the fundamental behavior of the mathematical functions offered by scipy.
However, as exposed by the test case in Figure~\ref{fig:example}, the output of the \code{differential\_entropy} function changes from 2.3588 to 2.5285 after the PR is applied.
That is, the PR causes an unintended behavioral difference, as the output of the function should not change for the same input data just because the internal handling of arrays changes.
The regression remained unnoticed by the developers, who merged the PR into the code base without realizing the unintended behavioral change.

\begin{figure}
	\begin{flushleft}
		\underline{PR title}:\\
		ENH: stats.differential\_entropy: add array API support

		\vspace{.5em}
		\underline{PR description}:\\
		...\\
		Adds array API support to differential\_entropy\\
		...

		\vspace{.5em}
		\underline{\name{}-generated test case}:
		\begin{lstlisting}
import numpy as np
from scipy.stats import differential_entropy

values = np.array([1, 1, 2, 3, 3, 4, 5, 5, 6, 7, 8, 9, 10, 11])
result = differential_entropy(values)
print(result)
		\end{lstlisting}

		\vspace{.5em}
		\underline{Output before PR}: 2.3588

		\vspace{.5em}
		\underline{Output after PR}: 2.5285

		\vspace{.5em}
		\underline{\name{}'s classification}:
		\emph{Unintended behavioral change}

		Differential entropy calculations should yield the same results for the same datasets unless there's a specific intention to change the underlying algorithm or methodology.
		The change in output appears unintended because the pull request does not specify altering the fundamental behavior or calculations of entropy estimators; it only indicates adding array API support.
\end{flushleft}
\caption{Motivating example.}
\label{fig:example}
\end{figure}

As illustrated by this example, regressions may easily remain unnoticed.
One reason is that, even in well tested software, the available test suite may not cover the code modified by a code change.
Automated regression test generation~\cite{Pacheco2007,Fraser2011a,icse2022-Nessie,Pizzorno2024,Ryan2024} can partially address this challenge, but lacks a useful test oracle:
If a regression test generator finds a test that exposes a behavioral difference between the code before and after a code change, it remains unclear whether this behavioral difference is intended.
A naive approach could report any behavioral difference as a regression, but this would lead to many false positives, because most code changes are supposed to change the behavior in some way, e.g., to fix a bug or to add a new feature~\cite{mostafa2017experience,icse2022-Nessie}.
 
This paper presents \name{}, an automated technique to detect regressions and other unintended behavioral changes by using natural language information associated with a code change as a test oracle.
Our key idea is to compare the intentions of a code change, as provided informally in natural language, with behavioral changes exposed by generated regression tests.
The approach checks PRs for unintended behavioral changes by performing three steps:
(1) At first, given the code diff associated with the PR, \name{} performs a targeted test generation aimed at finding tests that expose behavioral differences between the original and the modified code.
Motivated by the recent success of large language models (LLMs) in generating effective tests~\cite{Yuan2024,Yang2024,Pizzorno2024,Alshahwan2024,Ryan2024,Kang2023}, \name{} uses an LLM to generate tests that exercise the modified code.
(2) Next, the approach compares the behavior of the original and the modified code by executing the generated tests on both versions of the code.
(3) Finally, for any test that exposes a behavioral difference, \name{} classifies the difference as intended or unintended, and reports differences that are likely unintended to the user.
For the classification, \name{} also uses an LLM, which is provided with the title, description, and commit messages associated with the PR, and then prompted with multiple questions to determine whether the behavioral difference is intended.
To the best of our knowledge, our approach is the first to turn natural language information describing a code change into an oracle for validating behavioral differences.

We envision \name{} to be continuously applied to code changes, e.g., in the form of PRs, either while a change is under discussion or shortly after it has been merged into the code base.
In this setup, our approach provides several benefits:
(i) It detects regressions and other unintended behavioral changes that would otherwise remain unnoticed.
(ii) It provides a reproducible test case that exposes the problem, which can later be used to validate a fix and prevent the regression from reoccurring.
(iii) It avoids many false positives that a traditional regression testing approach would produce, because it does not simply report any behavioral change as a problem, but compares them to the developer-documented intentions.
(iv) It reports problems at a point in time when developers are most receptive to feedback, i.e., while or shortly after a developer is working on a code change.
(v) It produces natural language explanations for the detected regressions, which can help developers quickly assess potential problems.

Our evaluation applies \name{} to hundreds of real-world PRs from four popular and complex Python projects on GitHub.
We show that the approach can detect regressions that have remained unnoticed based on the existing test suite and any other tools run as part of the continuous integration pipelines of the target projects.
In total, we find 30 code changes with unintended behavioral changes, including \nbRegressions{} regression bugs and \nbCoincidentalFixes{} PRs that, despite having another intention, coincidentally fix a bug.
Out of 13 regressions reported to the developers, 11 have been confirmed and 9 have also been fixed at this point, with the remaining two issues still being open.
The classifier that predicts whether a behavioral change is intended is effective in practice, with a precision of \precision{} and a recall of \recall{}.
Finally, we find the costs of using \name{} to be acceptable for real-world deployment, with \minutesPerPR{} minutes to check a PR and LLM costs of only \$\dollarsPerPR{}, on average.

In summary, this paper contributes the following:
\begin{itemize}
	\item \emph{Idea}. We are the first to present the idea of using natural language artifacts associated with a code change as the basis for a regression testing oracle.
	\item \emph{Technique}. We present a novel technique for automatically detecting regressions and other unintended behavioral changes.
	\item \emph{Evidence}. We report the results of applying the approach to hundreds of real-world PRs, which shows its effectiveness and cost-efficiency. 
	\item \emph{Dataset and implementation}. We make our dataset, which is the first of its kind, and our implementation publicly available to foster future research.
\end{itemize}

\section{Approach}

\subsection{Problem Statement}

Before presenting our approach, let us define the problem we address.
\name{} aims to detect regressions in code changes that are introduced via PRs in a software project.
The input to the approach is a pull request $\mathit{pr}=(t,d,\Delta,m_c,m_d)$ that contains a title~$t$, a description~$d$ of the intended changes, a diff~$\Delta$ that shows the changes to the code, the commit messages~$m_c$ of the code changes, and any messages~$m_d$ exchanged among the developers while discussing the PR.
Given $pr$, the goal of \name{} is to determine whether the changes in $\Delta$ introduce a behavioral difference that does not align with the intention of the PR.
That is, our \name{} approach yields one of two possible outputs:
$$
\mathit{Testora}(pr) = \begin{cases}
	$(\text{``unintended''}, c, e)$ & \begin{array}{l} \text{if $\Delta$ causes behavioral} \\ \text{differences that are} \\ \text{inconsistent with $\mathit{pr}$} \end{array} \\
	\text{``intended''} & \begin{array}{l} \text{otherwise} \end{array} \\

\end{cases}
$$

In case the approach yields ``unintended'', it also provides a test case~$c$ that exposes the behavioral difference and a natural language explanation~$e$ that describes why the behavioral difference is not in line with the intentions of the PR.

For the motivating example in Figure~\ref{fig:example}, the input to \name{} includes the PR title and description, as shown in the upper part of the figure.
\name{} yields ``unintended'', along with the test case shown in the middle part of the figure and the explanation shown at the bottom.

\subsection{Overview}

\begin{figure}
	\centering
	\includegraphics[width=.8\linewidth]{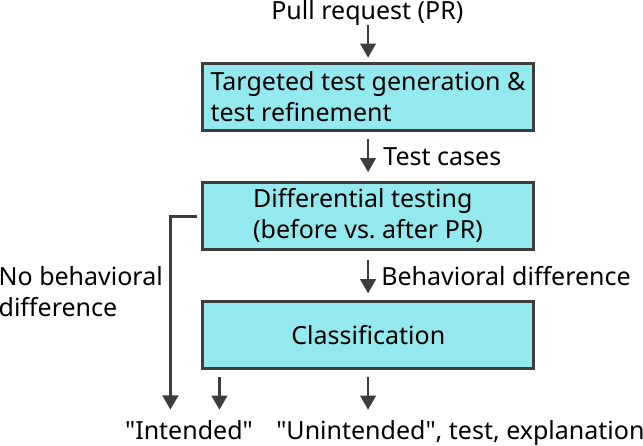}
	\caption{Overview of the approach.}
	\label{fig:overview}
\end{figure}

To address the problem defined above, \name{} follows the workflow shown in Figure~\ref{fig:overview}.
The approach consists of three main steps.
Given a PR, the first step is to generate tests aimed at exercising the modified code.
We introduce an LLM-based test generation and test refinement technique for this purpose (Section~\ref{sec:test generation}).
Based on the generated test cases, the next step compares the behavior of the original and the modified code via differential testing (Section~\ref{sec:diff testing}).
If the differential testing reveals a behavioral difference, the important question is whether this difference is intended, e.g., due to a new feature added by the PR, or unintended, e.g., due to a regression bug introduced by the PR.
To answer this question, \name{} uses an LLM-based classifier that compares the intention of the PR, as given in the PR title, description, and commit messages, with the observed behavioral difference (Section~\ref{sec:classifier}).
If the classifier determines that the behavioral difference is unintended, \name{} provides the test case that exposes the problem, along with a natural language explanation of why the behavior does not match the PR's intention.
The remainder of this section provides details on the different components of the approach.

\subsection{Selecting Code Changes to Check in Detail}
\label{sec:PR filtering}

Our approach is designed to automatically check a large number of PRs.
To keep the time and monetary costs manageable, we apply two filters to ignore PRs that are out of scope for our approach or that are unlikely to introduce unintended behavioral differences.

The first filter focuses on PRs that modify at least one file in the main source code of the target project, as opposed to modifying only other files, such as a README file, test cases, or configuration files.
To identify the main source code files, we consider files with specific extensions, i.e., \code{.py}, \code{.pyx}, \code{.c}, \code{.cpp}, and \code{.h},\footnote{We consider C/C++ files because some Python projects have performance-critical parts implemented in C/C++.} and that start with either ``src/$\langle$m$\rangle$'' or ``$\langle$m$\rangle$'', where $\langle$m$\rangle$ is the module name of the project.
Moreover, we exclude files with a path that contains ``test''.
If a PR does not modify any of the main source code files, we ignore it.
We also ignore PRs that modify more than three main source code files, as these PRs tend to be too complex to analyze for \name{}.

The second filter ignores documentation-only changes.
To this end, we parse both the old and the new version of all source code files modified by a PR into an AST and check if any changes beyond comments are made.
If a PR modifies only comments or other documentation, we ignore it.
Finally, we exclude PRs where the title starts with ``DOC'', which is commonly used to indicate documentation-only changes.

\subsection{Targeted Generation and Refinement of Regression Tests}
\label{sec:test generation}

The first main step of \name{} is to generate tests that exercise the modified code, aiming to expose behavioral differences between the original and modified code.
Given the recent success of large language models (LLMs) in generating code and tests~\cite{Chen2021,Yuan2024,Yang2024,Pizzorno2024,Alshahwan2024,Ryan2024,Kang2023}, we use a combination of lightweight static analysis and LLM prompting for this purpose.
The approach consists of three sub-steps: (1) gathering information about the code change, (2) querying an LLM to generate tests, and (3) finding undefined references and asking the LLM to resolve them.

\subsubsection{Gathering Information About the Code Change}

For an LLM to generate effective tests, it needs to understand the context of the code change.
To this end, \name{} gathers two kinds of information.
First, the approach extracts the names of all functions that are modified by the PR.
Providing these names is important for the LLM to understand which parts of the project to exercise in the generated tests.
To extract the function names, the approach uses a lightweight static analysis:
For each hunk in the code change, the approach parses the new version of the code file into an AST, finds the location of the changes in the AST, and identifies the inner-most function that encloses this location.
The approach then concatenates the module name of the modified file and the function name to get the fully qualified function name, which is important for the LLM to understand how to access the modified function.
If there is no surrounding function or if a hunk spans multiple functions, then the approach prompts the LLM without mentioning specific function names.

Second, the approach extracts the diff of the code change.
Specifically, we extract two versions of the diff.
On the one hand, we extract a \emph{full diff}, which contains all modifications made by the PR, including any edits of documentation.
On the other hand, the approach extracts a \emph{filtered diff}, which includes only modifications of files that belong to the main source code, as defined in Section~\ref{sec:PR filtering}.

\subsubsection{Querying an LLM to Generate Tests}

Given the gathered information about the code change, the approach queries an LLM to generate tests that exercise the modified code.
Our prompt includes the fully qualified names of modified functions and one of the two variants of the diff.
The prompt asks the LLM to generate usage examples that expose behavioral differences introduced by the provided diff.
We further specify that the usage examples must be self-contained Python code that uses only the publicly exposed APIs of the project and does not depend on non-deterministic behavior, such as random number generation.
To enable the approach to understand and compare the runtime behavior triggered by a usage example, we ask the LLM to include \code{print} statements for relevant values, such as the return value of an API function.
We ask to generate ten usage examples that cover normal usage scenarios and ten usage examples that focus on corner cases.
To increase the diversity of the generated tests, we query the LLM twice: once with the full diff and once with the filtered diff.
This is based on the observation that asking the LLM in slightly different ways produces different tests.

We clean and de-duplicate tests by removing any tests that invoke ``private'' functions.
The Python language does not have any notion of ``private'' functions, but most projects follow the convention to prefix functions not intended for client usage with an underscore.
Based on this convention, our approach discards tests that call such a function.
The rationale is that the tests should only exercise the public API of the project, as this is what users of the project interact with.
Finally, the approach discards duplicates, i.e., syntactically identical tests.

\subsubsection{Finding and Fixing Undefined References}

A common problem we observed in preliminary experiments is that some generated test cases refer to variables or functions that are not defined in the test case itself.
This problem is usually easy to fix, e.g., by adding an import statement or by defining a missing variable, but doing this manually for each test case would be too time-consuming.
Instead, we use a lightweight static analysis to find any undefined references, and then ask the LLM to refine the test case.
The prompt provides the initial test case, a list of identifiers that are referred to, but that are not properly defined, and then asks the LLM to provide a refined test case that resolves the undefined references.
Our refinement step is similar to conversational approaches to program repair~\cite{Xia2024a} and test assertion generation~\cite{Hayet2024a}, but uses a lightweight, AST-based static analysis, instead of executions, for feedback.

\subsection{Comparison of Behavior Before and After the Change}
\label{sec:diff testing}

Given the set of generated test cases, the next step of \name{} is to compare the behavior of the original and the modified code.
To this end, the approach starts by building the target project twice: at the commit just before the PR and at the commit with the PR applied.
Each commit is built in a containerized environment (based on Docker), providing two environments $\mathit{env}_{\mathit{old}}$ and $\mathit{env}_{\mathit{new}}$.
The motivation for using containers is to isolate these environments from each other and from the environment that \name{} itself is running in.

For each generated test case~$c$, the approach executes the test in both the old and the new environment.
These executions result in two outputs:
$o_\mathit{old} = \mathit{execute}(c, \mathit{env}_{\mathit{old}})$
and 
$o_\mathit{new} = \mathit{execute}(c, \mathit{env}_{\mathit{new}})$
The output from a test execution consists of all console output and any stack traces generated by exceptions.
Many of the LLM-generated tests contain \code{print} statements, which enables the approach to inspect the behavior of the code from its output.

Based on the two outputs $o_\mathit{old}$ and $o_\mathit{new}$, \name{} compares the behavior of the old and the new code.
If the outputs are exactly the same, i.e, $o_\mathit{old} = o_\mathit{new}$, then the test execution has not exposed any behavioral difference, and the approach moves on to the next test.
In contrast, if the outputs differ, i.e., $o_\mathit{old} \neq o_\mathit{new}$, then \name{} performs several steps aimed at focusing on relevant differences and at reducing the test case to a minimal example that still exposes the difference:
\begin{enumerate}
	\item Check whether both the old and the new version have failed with an exception.\footnote{Even if both versions raise the same kind of exception, the concrete outputs $o_\mathit{old}$ and $o_\mathit{new}$ often differ, e.g., because line numbers in the stack trace may differ.}
	If yes, we discard this test case because the test case likely raises a legitimate exception, e.g., caused by using the APIs of the target project incorrectly, which is not a regression.
	\item Validate the difference between $o_\mathit{old}$ and $o_\mathit{new}$ by re-executing the test with both the old and the new code. This step is to avoid flaky tests, which may cause non-deterministic differences in output.
	If the difference disappears when re-executing a test, the approach discards the test.
	\item Iteratively reduce the test case by removing one line at a time at the end and by checking if the outputs still differ.
	This process continues until no difference is found anymore or until the entire test case has been removed.
	The rationale for simply removing lines at the end, instead of using more sophisticated program reduction techniques~\cite{Zeller2002,misherghi2006hdd,ase2017-GTR}, is that tests are mostly linear programs without control flow statements.
	All further steps are performed on the reduced test case.
	\item Check if the difference is still present in the latest commit of the project, to avoid reporting regressions that have already been fixed.
	To this end, the approach builds the target project into a third environment $\mathit{env}_{\mathit{latest}}$ at the latest commit of the main branch, and re-executes the test case, which results in $o_{latest} = \mathit{execute}(c, \mathit{env}_{\mathit{latest}})$.
	If $o_{new} \neq o_{latest}$, then the developers have further changed the code affected by the PR after the PR had been applied.
	In this case, the approach discards the test case, as the observed behavioral difference is not relevant anymore.
	If \name{} is used continuously for every PR before it gets merged, this step can be skipped, as the approach finds regressions before they even get merged into the main branch.
\end{enumerate}

Any behavioral difference that remains after these steps is considered to be potentially unintended, and the tuple $(c, o_\mathit{old}, o_\mathit{new})$ is passed on to the final step of \name{} for classification.

\subsection{Classification of Behavioral Differences}
\label{sec:classifier}

\begin{figure}
	\begin{flushleft}
		\underline{PR title}:\\
		ENH: \code{special.logsumexp}: improve precision when one element is much bigger than the rest
	
		\vspace{.5em}
		\underline{PR description}:\\
		... can lose precision when one element is much bigger than the rest, especially when the exponential of it is close to 1. This improves the precision as described in the issue.
		...
		
		\vspace{.5em}
		\underline{\name{}-generated test case}:
		\begin{lstlisting}
import numpy as np
from scipy.special import logsumexp

a = np.array([1.0, 2.0, 3.0])
result = logsumexp(a)
print(result)
		\end{lstlisting}
		
		\vspace{.5em}
		\underline{Output before PR}: 3.4076059644443806
		
		\vspace{.5em}
		\underline{Output after PR}: 3.40760596444438
		
		\vspace{.5em}
		\underline{\name{}'s classification}:
		\emph{Intended behavioral change}
		
		While both outputs are very close and essentially represent the same value, the second output has fewer decimal places, which could be seen as a minor change in the representation of the result due to numerical precision improvements. 
		The change is intended by the developer, as the pull request specifically states that it aimed to improve precision for the \code{logsumexp} function when working with numbers that have a wide range.	
	\end{flushleft}
	\caption{Example of an intended behavioral change.}
	\label{fig:intended example}
	\end{figure}

The final step of our approach is to determine, for any behavioral difference found, whether the difference is intended or unintended.
This step involves two key challenges:
First, not every behavioral difference is a problem.
The reason is that many PRs are supposed to modify some behavior, e.g., to fix a bug or to add a new feature.
The challenge is to distinguish between intended and unintended differences.
For example, the behavioral difference shown in Figure~\ref{fig:example} is unintended, as the change in the output of the \code{differential\_entropy} function is not intended by the PR.
In contrast, consider the PR shown in Figure~\ref{fig:intended example}, which aims to improve the precision of the \code{logsumexp} function in the \code{scipy.special} module.
The PR also causes a behavioral difference, but unlike the previous example, the difference is intended, as the PR specifically aims to improve the precision of the function.
Second, some behavioral differences may not be directly intended by the PR, but at the same time, are so minor that they are irrelevant in practice.
Typical examples include minor differences in error messages, e.g., caused by different object addresses, or minor changes in formatting.

Our key insight is to address these challenges by providing an LLM with the natural language information provided in the PR, such as the PR title, description, and commit messages.
The idea is that the intention of the code change is often documented in these natural language artifacts, and that we can use this information to determine whether a behavioral difference is intended or unintended.
That is, we use the natural language information as an oracle for regression testing.

To classify whether a behavioral difference is intended, \name{} performs two sub-steps, (i) gathering relevant contextual information and (ii) a multi-question, LLM-based classifier, which we describe in the following.

\subsubsection{Gathering Relevant Context}

To enable the LLM classifier to take an informed decision, we provide it with relevant context about the code change.
The context includes the following information:
\begin{itemize}
	\item The name of the project.
	\item The fully qualified names of the functions that are modified by the PR.
	\item The title $t$ of the PR.
	\item The description $d$ of the PR.
	\item The diff $\Delta$ of the code change.
	\item The commit message(s) $m_{c}$ associated with the code diff.
	\item Any discussion comments $m_{d}$ associated with the PR.
	\item The test case $c$ that exposes the behavioral difference.
	\item The outputs $o_\mathit{old}$ and $o_\mathit{new}$ of the test execution with the old and the new code.
	\item The docstrings (if available) of all functions invoked in the test case.
\end{itemize}

To extract the PR-related information, we query the GitHub API.
The docstrings of functions are retrieved using the Language Server Protocol (LSP).
Retrieving docstrings is important for the LLM to understand the pre-conditions and the intended behavior of the functions invoked in the test case.

\subsubsection{Multi-Question, LLM-Based Classifier}
\label{sec:multi-question}

Based on the gathered context, \name{} prompts the LLM to determine whether a behavioral difference is intended or unintended.
One possible approach, which we call the \emph{single-question classifier}, would be to ask the LLM a single question, e.g., ``Is the behavioral difference intended?''.
In initial experiments, we found that the single-question classifier commonly gives suboptimal answers. 
For example, the classifier may report a behavioral difference as unintended, but the test case violates a pre-condition of the tested API, which implies that the behavioral difference is irrelevant in practice.

To address this issue, we present a \emph{multi-question classifier} that asks the LLM multiple questions designed to avoid common pitfalls of the single-question classifier.
\name{} prompts the LLM with the following five questions:
\begin{enumerate}
	\item Is the different output a noteworthy change in behavior, such as a completely different value being computed, or is it a minor change, such as a change in a warning/error message or a change in formatting?
	\item Is the different output likely due to non-determinism, e.g., because of random sampling or a non-deterministically ordered set?
	\item Does the usage example refer only to public APIs of the project, or does it use any project-internal functionality?
	\item Does the usage example pass inputs as intended by the API documentation, or does it pass any illegal (e.g., type-incorrect) inputs?
	\item Does the different output match the intent of the developer of the pull request?
\end{enumerate}

We instruct the LLM to answer the questions in a structured, JSON-based format.
For each question, the LLM is asked to first provide its thoughts, i.e., the reasoning behind the answer, and then to provide the actual answer.
The rationale for this structure is to force the LLM to reason about the question before giving an answer, which we found to improve the quality of the answers in preliminary experiments.

The answers received by the multi-question classifier are a 5-tuple $(a_1, a_2, a_3, a_4, a_5)$.
\name{} considers a behavioral difference as unintended if the LLM determines the difference to be noteworthy ($a_1$), deterministic ($a_2$), triggered via public APIs ($a_3$) using legal inputs ($a_4$), and causing an unintended difference in output ($a_5$).
Otherwise, i.e., if any of the five answers suggests the behavioral change to be expected or irrelevant, the approach considers the behavioral difference as intended.
If the approach concludes that the difference is unintended, it provides the test case~$c$ and the outputs $o_\mathit{old}$ and $o_\mathit{new}$ to the developer, along with the natural language thoughts~$e$ by the LLM explaining why the behavioral difference is unintended.

For example, the approach classifies the behavioral differences in Figures~\ref{fig:example} and~\ref{fig:intended example} as unintended and intended, respectively.
The bottom of the figures show the natural language explanations provided by the LLM.

\section{Evaluation}

Our evaluation applies \name{} to real-world Python projects to answer the following research questions (RQs):
\begin{itemize}
	\item \emph{RQ1: Effectiveness at finding real-world problems}. How effective is the approach at detecting regressions in real-world projects?
	\item \emph{RQ2: Effectiveness of test generation}. How effective is the approach at exercising changed code and at revealing behavioral differences?
	\item \emph{RQ3: Accuracy of classifier}. How accurate is the approach at distinguishing between intended and unintended behavioral differences?
	\item \emph{RQ4: Costs}. What are the computational costs of the approach?
\end{itemize}

\subsection{Experimental Setup}

\paragraph{Target projects}
Even though the approach is mostly language-agnostic, we focus on Python projects in our evaluation.
We select four real-world Python projects based on the following criteria:
\begin{itemize}
	\item Popularity: We discard projects with less than 5,000 stars on GitHub, aiming to target projects that are widely used and well maintained.
	\item Libraries: To ensure that the projects have a well-defined API that can be tested via generated tests we focus on libraries, i.e., we discard projects that are end-user applications, frameworks, and tutorial-style projects.
	\item PRs: We consider a project only if it has at least 1,000 closed PRs. This criterion ensures that the targeted projects follow a PR-based development process and have a sufficient number of PRs to evaluate the approach.
	\item Setup and requirements: Since the approach requires building and installing the target projects at different commits, we create a Docker container for each project. We discard projects where we were unable to create a container that reliably builds and installs the project.
	\item Diversity: To cover different application domains, we discard projects that are very similar to each other, such as multiple machine learning libraries.
\end{itemize}

\begin{table}
	\centering
	\setlength{\tabcolsep}{1pt}
	\caption{Projects used in the evaluation.}
	\label{tab:projects}
	\begin{tabular}{@{}llrr@{}}
		\toprule
		Name & Application domain & Stars & Closed PRs \\
		\midrule
		keras & Deep learning & 62,500 & 7,752 \\
		marshmallow & Serialization and deserialization  & 7,100 & 1,198 \\
		pandas & Data analysis and manipulation & 44,500 & 33,249 \\
		scipy & Scientific computing & 13,300 & 11,572 \\
		\bottomrule
	\end{tabular}
\end{table}

Table~\ref{tab:projects} provides an overview of the four selected projects.
The number of stars and closed PRs shown in the table are as of February 4th, 2025.
All four projects are mature, widely used in the Python ecosystem, and provide sufficient complexity to make automated regression testing challenging.

\paragraph{Pull requests}
\label{sec:PR datasets}
For RQ1, we apply the approach during several testing campaigns to batches of the most recent PRs of the target projects.
This process was repeated several times over the course of about six months. 
During a real-world deployment, \name{} would continuously analyze PRs as they are created, but for the evaluation in this paper, we analyze PRs in batches to reduce human time and computational costs.
Addressing RQ3 requires a labeled dataset, where each entry consists of a PR, a test case generated by \name{} that exposes a behavioral difference, and a label indicating whether the behavioral difference is intended or unintended.
We create such a dataset by manually inspecting a randomly sampled subset of the PRs considered in RQ1 where the approach found a behavioral difference.
The dataset consists of 164 labeled entries, of which 139 are intended and 25 are unintended behavioral changes.
%
For RQ2 and RQ4, we apply the approach to a systematically gathered range of up to 500 PR numbers per target project.
Specifically, we consider a range of consecutive PRs from each target project, starting from a PR number $n$, where $n$ was the first PR after a specific date (October 2, 2024).
For each $i$ in $n$ to $n+500$, we consider it as a target PR if $i$ is the number of a PR (and not of an issue, as issues use the same numbering scheme) and the PR was merged.
This process results in a total of 1,274 PRs across the four projects.

\paragraph{LLMs}
The approach interacts with an LLM to generate tests and to classify behavioral differences exposed by these tests.
As a default, we use GPT-4o-mini from OpenAI as the LLM.
This decision is based on the results of preliminary experiments that showed GPT-4o-mini to provide a good trade-off between effectiveness and costs.
For the important tasks of classifying behavioral differences, RQ3 also evaluates \name{} with the GPT-4o and DeepSeek-R1 models.

\paragraph{Implementation and hardware}

The approach is implemented in Python~3 in about 4,600 lines of non-empty, non-comment Python code.
We use the libCST library and the built-in Python AST module to parse and analyze the Python code of the target projects, and also to check for undefined references in generated test cases.
To interact with the GitHub API and the git repositories of the target projects, we use the PyGithub and GitPython libraries, respectively.
The implementation uses the Jedi Language Server via the multispy library~\cite{multilspy} to retrieve the docstrings of functions called in the generated tests.
The target projects are installed in separate Docker containers, and we use the Python docker library to interact with the containers.
In particular, all test executions are performed in these Docker containers to ensure that the tests run in a clean environment and do not interfere with the host system. 

All experiments are performed on a server with 48 Intel Xeon CPUs with 2.2GHz and 256GB of RAM running Ubuntu 22.04.
We run multiple experiments in parallel to reduce the time needed to process all PRs.
To distribute PR analysis tasks across multiple instances of the approach, we use a MySQL database to store the PRs to check and the results of \name{}.

\subsection{RQ1: Effectiveness at Finding Real-World Problems}

To validate the effectiveness of \name{} at finding problems in real-world projects, we apply it to PRs of the target projects.
When the approach detects a potential regression, we manually inspect the PR and the report generated by our approach.
Based on the inspection, each report by \name{} belongs to one of three possible outcomes: regression, coincidental fix, or false positive.
We discuss the first two cases in the following, and will evaluate false positives in RQ3.
Table~\ref{tab:success cases} provides an overview of the results.

\begin{table*}
	\centering
	\setlength{\tabcolsep}{2pt}
	\caption{Real-world regressions and coincidental fixes detected by \name{}.}
	\label{tab:success cases}
	\begin{tabular}{@{}rllp{27.5em}lll@{}}
		\toprule
		Id & Project & PR & Description & Kind & Issue & Status \\
		\midrule
		1 & keras & 19814 & Incorrect handling of \scode{NaN} in \scode{in\_top\_k} function in Jax backend & Regression & \anonID{19995} & Confirmed and fixed \\

		2 & marshmallow & 1399 & Refactoring coincidentally fixes incorrectly returned function arguments & Coincidental fix & -- & Nothing to report \\

		3 & marshmallow & 2698 & Incorrectly formatted timezone offset & Regression & -- & Fixed independently \\

		4 & marshmallow & 2699 & \scode{get\_fixed\_timezone} raises exception on floating point timezone offset & Regression & -- & Fixed independently \\

		5 & marshmallow & 2700& \scode{missing} argument of \scode{Field} class gets ignored & Regression & --  & Fixed independently \\

		6 & marshmallow & 2701 & Passing \scode{missing} argument to constructor of \scode{Field} class raises exception & Regression & -- & Fixed independently \\

		7 & pandas & 55108 & Intended to improve performance, but introduced bug that causes wrong output of \scode{Index.difference} function & Regression & \anonID{58971} & Confirmed and fixed \\
		
		8 & pandas & 56841 & Intended to improve performance, but introduced a bug that causes wrong output of \scode{Index.join} function & Regression & \anonID{58603} & Reported \\

		9 & pandas & 57034 & \scode{Series.combine\_first} produces wrong output for series containing \scode{None} & Regression & \anonID{58977} & Confirmed \\

		10 & pandas & 57046 & Intended to fix a bug in \scode{groupby.idxmin} related to extreme values, but also fixes bug triggered by \scode{NaN} & Coincidental fix & -- & Nothing to report \\

		11 & pandas & 57205 & Intended to improve performance, but also fixes a bug in handling \scode{None} values in \scode{DataFrame} constructors & Coincidental fix & -- & Nothing to report \\

		12 & pandas & 57399 & \scode{interval\_range} ignores type of \scode{start} parameter & Regression & \anonID{58964} & Reported \\

		13 & pandas & 57595 & Intended to improve performance, but also fixes a bug where \scode{DataFrame.groupby} returns an invalid value & Coincidental fix & -- & Nothing to report \\ 

		14 & pandas & 58376 & Intended to improve performance, but also fixes a bug related to calling \scode{RangeIndex.searchsorted} with a negative step & Coincidental fix & -- & Nothing to report \\ 

		15 & pandas & 60461 & Intended to improve performance, but also fixes wrong return value of \scode {construct\_1d\_object\_array\_from\_listlike} & Coincidental fix & -- & Nothing to report \\

		16 & pandas & 60483 & Backport of PR 60461 (see above) & Coincidental fix & -- & Nothing to report \\

		17 & pandas & 60538 & Adding \scode{DataFrames} with misaligned \scode{MultiIndex} produces \scode{NaN} despite \scode{fill\_value=0} & Regression & \anonID{60903} & Confirmed and fixed \\

		18 & scipy & 19263 & \scode{fft.hfftn} fails on list inputs & Regression & \anonID{21207} & Confirmed and fixed \\

		19 & scipy & 19428 & Should raise exception when \scode{stats.levene} is called with a single sample & Regression & -- & Fixed independently \\

		20 & scipy & 19680 & \scode{stats.shapiro} raises an error given lists of extreme integers & Regression & \anonID{21205} & Confirmed, won't fix \\

		21 & scipy & 19776 & Intended to vectorize computation, but also fixes bug triggered by passing \scode{inf} to \scode{stats.rankdata} & Coincidental fix & -- & Nothing to report \\

		22 & scipy & 19853 & Rewriting of internal functions coincidentally fixes bug that gave incorrect mean value of empty sparse matrix & Coincidental fix & -- & Nothing to report \\

		23 & scipy & 19861 & Bug fix also improves robustness of saving matrices into \scode{.mat} files & Coincidental fix & -- & Nothing to report \\

		24 & scipy & 20089 & \scode{special.hyp2f1} gives wrong result for extreme inputs & Regression & \anonID{20988} & Confirmed and fixed \\

		25 & scipy & 20751 & Value returned by \scode{stats.bartlett} is negative when exact value would be zero & Regression & \anonID{21152} & Confirmed and fixed \\

		26 & scipy & 20974 & \scode{stats.combine\_pvalues} gives result with wrong dimensionality & Regression & \anonID{21106} & Confirmed and fixed \\

		27 & scipy & 21036 & Intended to add array API support, but also fixes a bug that caused \scode{stats.tsem} to incorrectly return \scode{inf} & Coincidental fix & -- & Nothing to report \\

		28 & scipy & 21076 & \scode{stats.differential\_entropy} with integer dtype gives incorrect result & Regression & \anonID{21192} & Confirmed and fixed \\

		29 & scipy & 21553 & Performance improvement leads to incorrect output of \scode{expm} & Regression & -- & Fixed independently \\

		30 & scipy & 21768 & Slicing sparse matrix with \scode{None} gives result different from numpy's matrices & Regression & \anonID{22458} & Confirmed and fixed \\

		\bottomrule
	\end{tabular}
\end{table*}

\subsubsection{Regressions}
We categorize a report produced by \name{} as a \emph{regression} if the PR introduces a bug that has remained unnoticed by the developers by the time of merging the PR.
In this case, we check whether the problem still exists in the latest commit of the project, and if yes, report the problem to the developers.
In total, the approach has detected \nbRegressions{} regressions, out of which 13 had still been present in the latest commit of the project by the time that we inspected the PRs.
We reported these 13 regressions to the developers.
As of July 2025, 11/13 of them have been confirmed, and 9 out of the 11 confirmed problems have been fixed in reaction to our report.
2/13 of the reported regression bugs are still open.
1/13 issue has been confirmed as a bug but eventually closed as ``won't fix'' by the developers because the problem is unlikely to occur in practice while fixing it would affect a lot of code.
The remaining 6/19 PRs had already been fixed independently by the developers by the time that we checked the PRs.
If a technique like \name{} had been used continuously during development, all 19 regressions could have been detected earlier and fixed before the PR was even merged into the code base.

All detected regressions were missed by the regression test suites that are executed as part of the continuous integration pipelines of the studied projects. This is despite the fact that the existing tests are very comprehensive. For example, for scipy (where Testora finds most regressions), executing all tests takes multiple CPU hours.

\begin{figure}[t]
	\begin{lstlisting}
import jax.numpy as jnp
from numpy import nan
from keras.src.backend.jax.math import in_top_k

r = in_top_k(targets=jnp.array([1, 0]),
    predictions=jnp.array([[.1, nan, .5], [.3, .2, .5]]), k=2)
print(r)  # [False  True] vs. [True  True]
	\end{lstlisting}
	\caption{Regression in keras (\#1 in Table~\ref{tab:success cases}), leading to incorrect output.}
	\label{fig:regression diff output keras}
\end{figure}

\begin{figure}[t]
	\begin{lstlisting}
from scipy.fft import hfftn

x = [[[1.0, 1.0], [1.0, 1.0]], [[1.0, 1.0],
      [1.0, 1.0]], [[1.0, 1.0], [1.0, 1.0]]]
hfftn(x)  # no exception vs. exception
	\end{lstlisting}
	\caption{Regression in scipy (\#18 in Table~\ref{tab:success cases}), leading to an exception.}
	\label{fig:regression exception scipy}
\end{figure}

\paragraph{Examples}

Beyond the example in Figure~\ref{fig:example}, we provide additional examples of regressions detected by \name{}.
Figure~\ref{fig:regression diff output keras} shows a test case that exposes a regression bug in the keras project.
The PR, entitled ``Faster \code{in\_top\_k} implementation for Jax backend'', intends to improve performance for the Jax backend of keras.
Unfortunately, the optimized code incorrectly considers \code{nan} to be a large probability, which causes the \code{in\_top\_k} function to return incorrect results.
Figure~\ref{fig:regression exception scipy} shows a regression bug in scipy, which causes the \code{fft.hfftn} function to raise an exception when called with a list argument.
The PR that causes the bug intends to introduce GPU support for specific computations, which unfortunately, breaks some of the existing functionality.

\subsubsection{Coincidental Fixes}
In addition to regressions, \name{} also detects PRs intended to improve the code in some way, but that -- unknowingly to the developers -- also fix a bug.
We call such cases \emph{coincidental fixes}.
Overall, \name{} detects \nbCoincidentalFixes{} coincidental fixes, as listed in Table~\ref{tab:success cases}.
Because the bug is fixed by the PR, we generally do not report the problem to the developers.
However, we document coincidental fixes in our evaluation to show that the approach can also find bugs that have remained unnoticed by the developers.
Project maintainers using a technique like \name{} could benefit from knowing about coincidental fixes, e.g., to decide which PRs to backport to older releases or to help with documenting fixed bugs in the release notes of the project.

\paragraph{Examples}

\begin{figure}[t]
	\begin{lstlisting}
import pandas as pd

df = pd.DataFrame({'A': ['foo', 'bar', None], 'B': [1, 2, 3]})
grouped = df.groupby('A', dropna=False)
print(len(grouped))  # exception vs. 3
	\end{lstlisting}
	\caption{Bug in pandas that was coincidentally fixed by a PR meant to improve performance (\#13 in Table~\ref{tab:success cases}).}
	\label{fig:coincidental fix pandas}
\end{figure}

Figure~\ref{fig:coincidental fix pandas} shows a coincidental fix in the pandas project.
The PR intends to improve performance of the \code{DataFrame.groupby} function.
However, the revised code also fixes a bug that causes the function to raise an exception when grouping by a column that contains \code{None} values.

Another example (\#2 in Table~\ref{tab:success cases}) is a PR in the marshmallow project, which is described as ``minor refactor'', but coincidentally fixes an incorrect output of a utility function that returns the arguments of a given callable.

\subsection{RQ2: Effectiveness of Test Generation}

\begin{table}[t]
	\caption{Pull requests analyzed for RQ2 and RQ4.}
	\label{tab:prs}
	\setlength{\tabcolsep}{6pt}
	\begin{tabular}{@{}lrrrr@{}}
		\toprule
		Project & \multicolumn{4}{c}{Pull requests} \\
		\cmidrule{2-5}
		& Total & Ignored & Checked & Behavioral  \\
		&&&& difference found \\
		\midrule
		keras & 271 & 111 & 160 & 0 \\
		marshmallow & 138 & 73 & 65 & 8 \\
		pandas & 439 & 296 & 143 & 35 \\
		scipy & 426 & 269 & 157 & 57 \\
		\midrule
		Total & 1,274 & 749 & 525 & 100 \\
		\bottomrule
	\end{tabular}
\end{table}

In addition to evaluating the end-to-end effectiveness of \name{} in RQ1, the following two research questions evaluate the two main components of the approach in more detail.
RQ2 focuses on the effectiveness of the test generation component. 
To this end, we apply \name{} to the 1,274 PRs listed in the ``Total'' column of Table~\ref{tab:prs}.
As described in Section~\ref{sec:PR filtering}, \name{} ignores some PRs, e.g., PRs unlikely to introduce regressions, such as changes to test files or documentation-only changes.
This filtering affects 749/1,274 PRs, as shown in the ``Ignored'' column of Table~\ref{tab:prs}, leaving 525 PRs for \name{} to analyze in detail.
For the 525 PRs that \name{} analyzes in detail, we evaluate the effectiveness of the test generation component.
As shown in the last column of Table~\ref{tab:prs}, the approach finds a behavioral difference in 100/525 PRs, i.e., in 19\% of the PRs.

\begin{table}[t]
	\caption{Test generation results (min/avg/max per PRs).}
	\label{tab:testgen}
	\setlength{\tabcolsep}{2pt}
	\begin{tabular}{@{}lcc|cc@{}}
		\toprule
		Project & \multicolumn{2}{c}{Tests} & \multicolumn{2}{c}{Test executions} \\
		\cmidrule{2-3}
		\cmidrule{4-5}
		& Generated & With diff.\ cov.  & Total & Non-failing  \\
		\midrule
		keras & 2/27/40 & 0/14/40 & 4/76/160 & 0/48/156 \\
		marshmallow & 9/32/40 & 0/26/40 & 18/96/530 & 0/83/528 \\
		pandas & 2/31/40 & 0/18/40 & 4/107/1098 & 0/88/1059 \\
		scipy & 1/30/40 & 0/18/40 & 2/135/1481 & 0/101/1073 \\
		\bottomrule
	\end{tabular}
\end{table}

Table~\ref{tab:testgen} provides more detailed results of the test generation and testing process.
The table show the minimum/average/maximum values per PR, across the 525 checked PRs.
At first, we consider the number of generated tests.
In principle, the combined answers of the LLM consist of $2 \times 2 \times 10 = 40$ tests (Section~\ref{sec:test generation}).
In practice, there may be fewer tests because the LLM occasionally provides fewer tests than requested and because we de-duplicate tests.
During our experiments, we find that the approach generates between 27 and 32 unique test cases, on average per PR.

Next, we study whether the generated tests cover the code changed by the PR, which is a prerequisite for finding behavioral differences.
We consider a generated test to ``have diff coverage'' if it covers at least one line of the changed code.
As shown in Table~\ref{tab:testgen}, between 14 and 26 tests, on average per PR, have diff coverage.
That is, the majority of the generated tests (52\%--81\%) are successful at exercising changed code.

Finally, the last two blocks of Table~\ref{tab:testgen} show the number of test executions performed by \name{} when analyzing a single PR.
These numbers are higher than the number of generated tests because the approach re-executes tests while reducing tests that expose a behavioral difference and to filter flaky tests by validating each behavioral difference.
Overall, the approach executes between 76 and 135 tests, on average per PR.
The majority of these tests (63\%--86\%) are non-failing, i.e., they do not raise an exception.

\subsection{RQ3: Accuracy of Classifier}

Besides the test generation component, as evaluated in RQ2, the classifier that distinguishes between intended and unintended behavioral differences is crucial for the effectiveness of \name{}.
The following evaluates the accuracy of the classifier using different LLMs and prompting techniques.
We apply the classifier to the labeled dataset described in Section~\ref{sec:PR datasets}.
As the dataset is imbalanced in the number of intended and unintended behavioral differences, we evaluate the accuracy of the classifier by measuring precision, recall, and F1 score.
Precision here means the percentage of behavioral differences classified as unintended that are actually unintended.
Recall means the percentage of unintended behavioral differences that the approach recognizes as unintended.
The F1 score is the harmonic mean of precision and recall.

We evaluate the accuracy of the classifier using different LLMs and prompts.
As LLMs, we consider two state-of-the-art commercial LLMs, GPT-4o and its more economic variant, GPT-4o-mini, as well as a state-of-the-art open-source LLM, DeepSeek-R1 (671B).
We evaluate the classifier using two different prompts: the multi-question classifier described in Section~\ref{sec:multi-question} and a simpler, single-question classifier that asks the LLM only whether the behavioral difference is intended by the PR.

\begin{table}[t]
	\caption{Accuracy of classifier (bold = default configuration).}
	\label{tab:classifier}
	\setlength{\tabcolsep}{4pt}
	\begin{tabular}{@{}lrrr|rrr@{}}
		\toprule
		LLM & \multicolumn{3}{c}{Single-question} & \multicolumn{3}{c}{Multi-question} \\
		& \multicolumn{3}{c}{classifier} & \multicolumn{3}{c}{classifier} \\
		\cmidrule{2-4}
		\cmidrule{5-7}
		& Precision & Recall & F1 & Precision & Recall & F1 \\
		\midrule
		GPT-4o-mini & 49\% & 80\% & 61\% & \textbf{55\%} & \textbf{67\%} & \textbf{60\%} \\
		GPT-4o & 80\% & 64\% & 71\% & 71\% & 42\% & 53\% \\
		DeepSeek-R1 & 69\% & 36\% & 47\% & 83\% & 42\% & 56\% \\
		\bottomrule
	\end{tabular}
\end{table}

Table~\ref{tab:classifier} shows the accuracy of the classifier using the different LLMs and prompts.
We make three observations.
First, and perhaps most importantly, the results are relatively stable across different models and different variants of the classifier.
That is, the approach does not depend on a specific LLM or prompting technique, but generalizes well to the considered (and probably also future) models and prompts.
Second, no clear winner emerges among the considered LLMs:
While the overall best configuration, in terms of F1-score, is GPT-4o with the single-question classifier, both GPT-4o-mini and DeepSeek-R1 outperform GPT-4o when using the multi-question classifier.
To keep the costs of using \name{} manageable, we use GPT-4o-mini as the default model.
Third, when comparing the two prompting techniques, we again see a diverse set of results:
The single-question classifier outperforms the multi-question classifier in terms of F1-score for GPT-4o, but the multi-question classifier outperforms the single-question classifier for DeepSeek-R1.
For our default model, GPT-4o-mini, the multi-question classifier offers approximately the same F1 score as the single-question classifier (60\% vs.\ 61\%), but offers a better balance between precision and recall, which is why we use the multi-question classifier as the default in \name{}.

\subsection{RQ4: Costs}

\begin{figure}
	\centering
	\includegraphics[width=\linewidth]{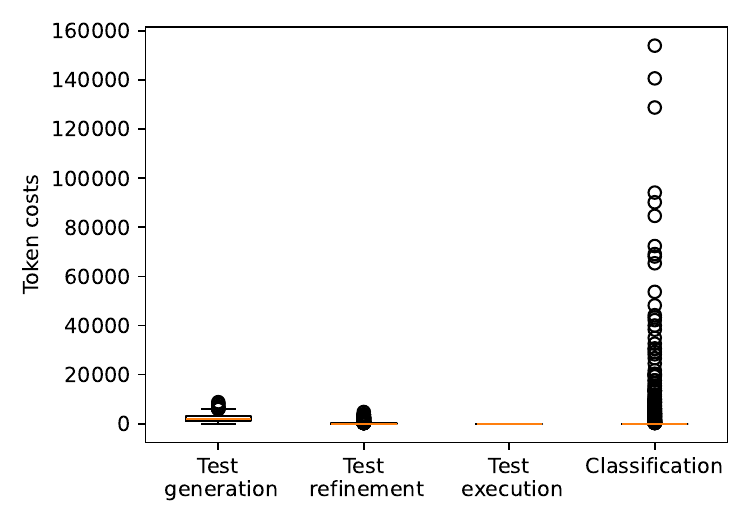}
	\caption{LLM tokens used per PR.}
	\label{fig:token_costs}
\end{figure}

Our final research question evaluates the computational costs imposed by \name{}.
We consider two types of costs: monetary costs imposed by LLM queries, and the time taken by the approach.
The monetary costs are calculated based on the number of input and output tokens used by the LLM.
On average per checked PR, the approach consumes \totalTokensPerPR{} tokens, of which \inputTokensPerPR{} are input tokens and \outputTokensPerPR{} are output tokens.
Figure~\ref{fig:token_costs} shows a breakdown of the tokens consumed during different steps of \name{}.
We find that test generation and classification consumes most tokens, whereas test refinement has negligible costs, and test execution (by design) does not consume any tokens.
The tokens used by the classifier show several outliers, which are due to long diffs and long discussions of PRs.
Yet, despite these outliers, the overall costs remain manageable.
Based on the pricing of OpenAI's GPT-4o-mini model as of February 17, 2025, the total token consumption results in monetary costs of \$\dollarsPerPR{} per checked PR.
Given the high human effort caused by regressions that remain unnoticed, we consider this cost acceptable for real-world deployment at a large scale.

\begin{figure}
	\centering
	\includegraphics[width=\linewidth]{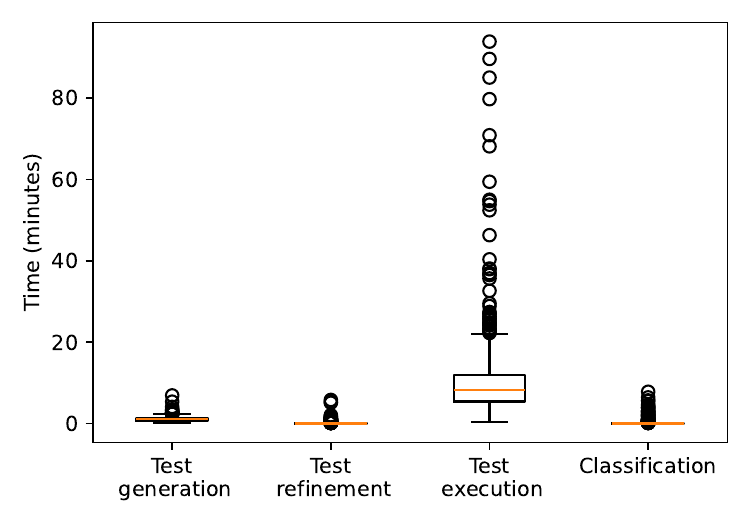}
	\caption{Time taken per PR.}
	\label{fig:time_costs}
\end{figure}

Besides LLM costs, we also evaluate the time taken by the approach.
On average per checked PR, the approach takes \minutesPerPR{} minutes to analyze a PR.
As shown in Figure~\ref{fig:time_costs}, the time is mostly spent on test execution.
A relatively large fraction of this time is for compiling the target project at a specific commit, which is a prerequisite for running the generated tests.
To put the time of \minutesPerPR{} minutes in perspective, we compare it to the time taken by the continuous integration (CI) effort performed by an average non-free repository that uses GitHub workflows~\cite{icse2024-workflows}: \name{}'s time corresponds to 39.5\% of the CI time spent when creating a new pull request and to 43.3\% of the CI time spent each time a developer pushes new commits into a repository.
While this was not our focus, we believe that the time could be further reduced, e.g., by pre-computing Docker images compiled at specific commits and by parallelizing the execution of different generated tests.

\section{Discussion}

\subsection{Limitations}

\name{} relies on automated test generation, which is most suitable for projects that have a well-defined API amenable to unit-level testing.
Extending our idea to projects with a more complex interface, e.g., a graphical user interface, is left for future work.
Another limitation is that very complex code changes, e.g., where the diff spans many files, are likely to be challenging for the approach.

\subsection{Threats to Validity}

Our evaluation is based on a set of four projects, which may not be representative of all software projects.
We mitigate this threat by selecting projects from different domains and by evaluating the approach on hundreds of PRs.
Another threat is that different LLMs may provide different results, which we mitigate by evaluating the classifier with three different LLMs.
The manual labeling to establish a ground truth for RQ3 is also a potential threat to validity, as it may be biased.
To mitigate this threat, the annotator is a senior researcher, who made two passes over all 164 examples: one pass to assign an initial label, and another pass to double-check all labels and ensure consistency across all examples.
Another potential bias is that the examples to label are randomly sampled from the set of PRs where \name{} found a behavioral difference, and hence, some projects are represented more frequently than others.
Finally, our results are limited to Python projects, and while the general idea of \name{} is likely to extend beyond Python, we do not have empirical evidence for this.

\section{Related Work}

\paragraph*{Test generation and regression testing}
To complement manually written tests, there are various approaches for automated test generation, including feedback-directed, random test generation~\cite{Pacheco2007}, symbolic and concolic execution~\cite{King1976,Sen2005,Cadar2008}, and search-based test generation~\cite{Fraser2011a}.
More recently, the community is focusing increasingly on LLM-based test generation, e.g., in combination with a search-based approach~\cite{Lemieux2023}, for fuzzing libraries~\cite{Deng2023} and entire applications~\cite{icse2024-Fuzz4All}, for bug reproduction~\cite{Kang2023}, for unit-level test generation~\cite{Yuan2024,Pizzorno2024,Ryan2024,Yang2024}, and for test augmentation~\cite{Alshahwan2024}.
In principle, these approaches could be integrated into the first step of \name{}, but our approach differs by generating tests that target a specific code change, instead of trying to increase overall coverage.
Our work also relates to work on selecting and prioritizing regression test cases~\cite{Harrold2001,Elbaum2002,Yoo2012}, but differs by generating new tests for the given code change.

\paragraph*{Test oracle problem}
Effective testing requires not only suitable test inputs, but also a way to identify unexpected behavior, also known as the test oracle problem~\cite{Barr2015}.
One line of work to address this problem infers oracles from API documentation, e.g., in the form of exception oracles~\cite{Goffi2016} or metamorphic relations~\cite{Blasi2021}, or based on regular expressions~\cite{Motwani2019}.
Another line of work uses deep learning and LLMs to generate likely oracles for a given test prefix~\cite{Watson2020,Dinella2022,Nie2023,Hayet2024a,Hossain2025}.
Unlike these approaches, \name{} focuses on code changes and proposes the novel idea of exploiting natural language associated with a code change as a basis for a regression test oracle.

\paragraph*{Reasoning about code changes and code differences}
Differential testing~\cite{McKeeman1998} compares the behavior of multiple implementations and has been applied, e.g., to compilers~\cite{Barany2018}, debuggers~\cite{fse2018}, symbolic execution engines~\cite{Kapus2017}, and quantum computing platforms~\cite{Wang2021b}.
Our work differs by comparing two versions of the same code, instead of two independently created implementations.
To compare two versions of the same code, prior work has explored change-oriented symbolic execution~\cite{Marinescu2013} and validating code changes via learning-guided execution~\cite{fse2023-LExecutor,fse2025-ChangeGuard}.
These approaches share the idea of exposing a behavioral difference between an older and a newer version of the code, but unlike \name{}, do not address the problem of determining whether a detected difference is intended.
Just-in-time defect prediction tries to predict the general likelihood that a commit introduces a bug~\cite{Hoang2019,Tabassum2020,zeng2021deep,morita2024tracejit}.
Our approach goes further by comparing the behavioral change to the developer's intent, as expressed in the PR description, and by providing a concrete test case that exposes a likely regression.
The TraceJIT approach~\cite{morita2024tracejit} is particularly related due to its dynamic features.
A key difference from our work is that \name{} is not based on generic features (e.g., the extent to which a dynamic trace changes), but instead checks any behavioral differences against the developer-expressed intent of the PR.
Work on untangling commits~\cite{Partachi2020a,li2022utango} is related in that it could serve as a pre-processing step before running \name{}, enabling our approach to handle more complex, entangled code changes.
The DCI approach~\cite{danglot2020approach} also uses tests to reason about behavioral changes induced by a code change, but unlike \name{}, assumes that a test covering the code change already exists.
Once a bug is known, techniques for retrospectively determining the bug-inducing commit have been proposed~\cite{sliwerski2005changes,An2023}.
Finally, other approaches help reason about code changes without specifically targeting regressions, e.g., by augmenting diffs with runtime information~\cite{Etemadi2023} and via a unified data representation of code changes~\cite{Wu2021}.

\paragraph*{Natural language and code}
Natural language embedded in code has been used to detect bugs~\cite{oopsla2018-DeepBugs} and name inconsistencies~\cite{Nguyen2020}, to infer type annotations~\cite{Hellendoorn2018,icse2019,fse2020}, and to predict formal specifications of code~\cite{Zhai2020a,Endres2024}.
Other work summarizes code into comments~\cite{Gros2020,Ahmad2020,Zhang2020,Mu2023}, detects inconsistencies between code and comments~\cite{Panthaplackel2021,Rong2025}, automatically updates comments when code changes~\cite{Panthaplackel2020a}, and propagates comments across related code elements~\cite{Zhai2020}.
Summarizing the likely intent of code into natural language can also support automated issue fixing~\cite{Ruan2024}.
Others have proposed to detect potential bugs by generating alternative implementations from a natural language specification~\cite{Li2023c,Liu2024}.
Like this work, the above techniques share the observation that natural language associated with code provides a rich resource.
Unlike prior work, we leverage this resource as the basis for a regression oracle and integrate it into an automatic regression testing technique.

\section{Conclusion}

Regression testing is a crucial part of software development, but it is challenged by the need to identify unintended behavioral changes.
We present \name{}, an approach that (i) generates regression tests targeted at code changed by a PR, (ii) executes these tests to identify behavioral differences introduced by the PR, and (iii) classifies the behavioral differences exposed by the tests as intended or unintended.
A key contribution of \name{} is to use natural language information associated with a code change as a basis for a regression testing oracle.
Our evaluation on real-world projects shows that \name{} is effective at finding 19 real-world problems, while imposing acceptable costs of \minutesPerPR{} minutes of computational time and \$\dollarsPerPR{} of monetary costs per checked PR.
We envision our approach to serve as a useful tool for developers to continuously and early on detect regressions and coincidental bug fixes.
Initial feedback from developers is very positive.
For example, the developers of scipy asked us to apply \name{} to more PRs and are interested in applying \name{} to their project in the future.

\section*{Data Availability}
Code and data associated with this work are available:\\
\url{https://github.com/michaelpradel/Testora}

\section*{Acknowledgments}

This work was supported by the European Research Council (ERC, grant agreements 851895 and 101155832) and by the German Research Foundation within the DeMoCo and QPTest projects.
The author also thanks Koushik Sen and Miryung Kim for hosting him during his sabbatical at UC Berkeley und UCLA, respectively, during which parts of this work were conducted.

\bibliographystyle{ACM-Reference-Format}
\bibliography{referencesMP,moreRefs}
\end{document}